\pdfoutput=1

\documentclass[12PT]{iopart}
\usepackage[pdftex]{graphicx}

\begin{document}

\title{HEAVY-QUARKONIA IN THE STAR EXPERIMENT}

\author{\footnotesize MAURO R. COSENTINO for the STAR COLLABORATION}

\address{Instituto de F\'{i}sica, University of S\~ao Paulo, Rua do Mat\~ao, 187 - travessa R S\~ao Paulo, SP 05508-090,Brazil}

\ead{mcosent@dfn.if.usp.br}

%\pacs{13.85.Qk, 13.20.Fc, 13.20.He, 25.75.Dw}
%\maketitle

\begin{abstract}

Heavy Quarkonium states modifications in relativistic heavy ion collisions have been 
of great interest since the proposal by Matsui and Satz of $J/\psi$ suppression as a signature of 
Quark-Gluon Plasma (QGP) formation. Recent studies suggest that the excited states 
$\chi_c$,  $\psi$(2S) and $\Upsilon$(3S) melt sequentially\cite{1,2} and the amount of observed suppression depends
on the state and medium conditions. Therefore, this suppression pattern may be used as a 
probe of the medium temperature. In this work we present preliminary results on the
charmonium and bottomnium measurements performed by the STAR experiment at RHIC for 
p+p and Cu+Cu collisions at $\sqrt{s_{NN}}=$200GeV

\end{abstract}

\section{Introduction}

The main ideia of this work is to make a comparation between the $J/\psi$ and $\Upsilon$ (1s+2s+3s) production in 
p+p and Cu+Cu. These heavy quarkonia states are measured in STAR through the $e^+e^-$ decay channel. From lattice
QCD calculations\cite{1,2} it is expected a suppression of these states in heavy ion collisions due to the sequential
melting of their excited states

\section{Experimental Setup}

The experimental setup for this measurement relies in basically three subsystems of the STAR experiment,\cite{3,4,5}
which are the Time Projection Chamber (TPC), Barrel Electromagnetic Calorimeter (BEMC) and Central Trigger Barrel (CTB),
this last one only for triggering $J/\psi$ events in p+p. All of these subsystems have full azimuthal coverage.
The TPC has $\vert\eta\vert<$1.8 and provides up to 45 ionization points to charged tracks within its acceptance, allowing $dE/dx$
and momentum reconstruction. The BEMC is a lead-scintillator sampling electromagnetic calorimeter with equal volumes 
of lead and scintillator, divided in 4800 towers within the $\vert\eta\vert<$1 range, each of them with a face area of 
$\Delta\phi,\Delta\eta=$(0.05,0.05). The detector resolution on deposited electromagnetic energy is 
$\frac{dE}{E}\sim\frac{16\%}{\sqrt{E}}$. TPC was fully installed for both, p+p and Cu+Cu runs, while the BEMC was
$\sim$3/4 installed for Cu+Cu and fully installed for p+p. The CTB is a barrel made if scintilator slats that are
sensitive to charged particles only.
As the heavy quarkonia states have really low cross-sections and their decay channel observed in STAR has a also
relatively low branching ratio, the measurements have to improve its statistics relying in either, a large minimum bias data 
set (MBT), which was the case for Cu+Cu, or in special dedicated triggers as in p+p wihich are ndetailed in the following section.

\section{Trigger Setup}

The STAR Quarkonia Trigger is a two level trigger system designed to optimize the STAR measurement capabilities of
the di-electron decay channel for the heavy quarkonia states. The following sub-sections details the trigger for 
J$\psi$, while the $\Upsilon$ trigger system is detailed in reference.\cite{6} For the Cu+Cu run there were not specific
quarkonia triggers, and the measurements rely on large MB ($\sim$40M events for $J/\psi$)  and a high tower trigger (HTT) 
($\sim$9M events for $\Upsilon$) data sets. This HTT trigger have similar effect as the L0-$\Upsilon$ trigger.\cite{6}
  The L0-$J/\psi$ trigger is a topological trigger where the decision is make at the hardware level. It divides the 
BEMC into 6 sectors in $\phi$ and 2 in $\eta$, called patches. There are required at least 2 BEMC towers in 
non-adjacent patches, with energy above the threshold of 1.2 GeV (high-towers, or HT). With all these requirements 
fulfilled the L2-$J/\psi$ is started.
  The L2 is a software level decision. It is basically the same algorithm described in [6], but with the extra
requirement that the HT that seeds the L2 must match a CTB slat adc$>$3 in order to avoid photons. The remaining
difference is the value of the invariant mass parameter, according to the specific quarkonium state.

\section{Electron Identification}

The electron identification\cite{6} is a central issue in the STAR Quarkonium Program once its measurements come from the
$e^+e^-$ decay channel. The problem is to separate electrons from hadron contamination, dominated mainly by pions. The 
electron identification procedure for quarkonia analysis is basically to choose TPC tracks with momentum $p>$1 GeV/c 
and 3$<dE/dx<$4.6. After this first selection the track is extrapolated to its correspondent BEMC tower, its energy 
$E_{tow}$ is obtained and the ratio $p/E_{tow}$ is computed. Those with $p/E_{tow}<$2 are selected as electrons(positrons).

\section{Results}

Once the electrons (positrons) are selected they are put together in pairs and have their invariant mass calculated by 
the expression
    \begin{eqnarray}
      M^2 = 2E_1E_2(1-cos\theta_{12})\,
      \label{eq:mass2}
    \end{eqnarray}
where $p_i$ is the momentum of the $i-$th particle and $\theta_{12}$ is the angle between them. To estimate the background 
a mass spectrum is built up from the geometric mean between the positvely charged pairs with the negatively charged ones 
multiplied by 2. The net signals (unlike signed pairs $-$ like signed) are presented in figure 1. From there it is 
possible to see the enhancement of the signals due to the trigger system by comparing the different significances for
p+p and Cu+Cu. In the $J/\psi$ case it increases from $\sim$2.5 in Cu+Cu to $\sim$5.0 in p+p, and for $\Upsilon$ 
from $\sim$2.6 in Cu+Cu to 3.0$\sigma$ in p+p. The p+p $\Upsilon$ measurement cross section calculated
is $BR\times\frac{d\sigma}{dy}\vert_{y=0}=$91$\pm$28$({\it stat.})\pm$22$({\it sys.)}$  pb  (BR accounts for branching
of 1S+2S+3S states). This result agrees very well with pQCD-CEM calculations\cite{7,8} presented in figure 2. For the
other quarkonium states further analysis are needed before we can quote their cross sections.
\begin{figure}[th]
\begin{center}
\includegraphics[width=1.0\textwidth]{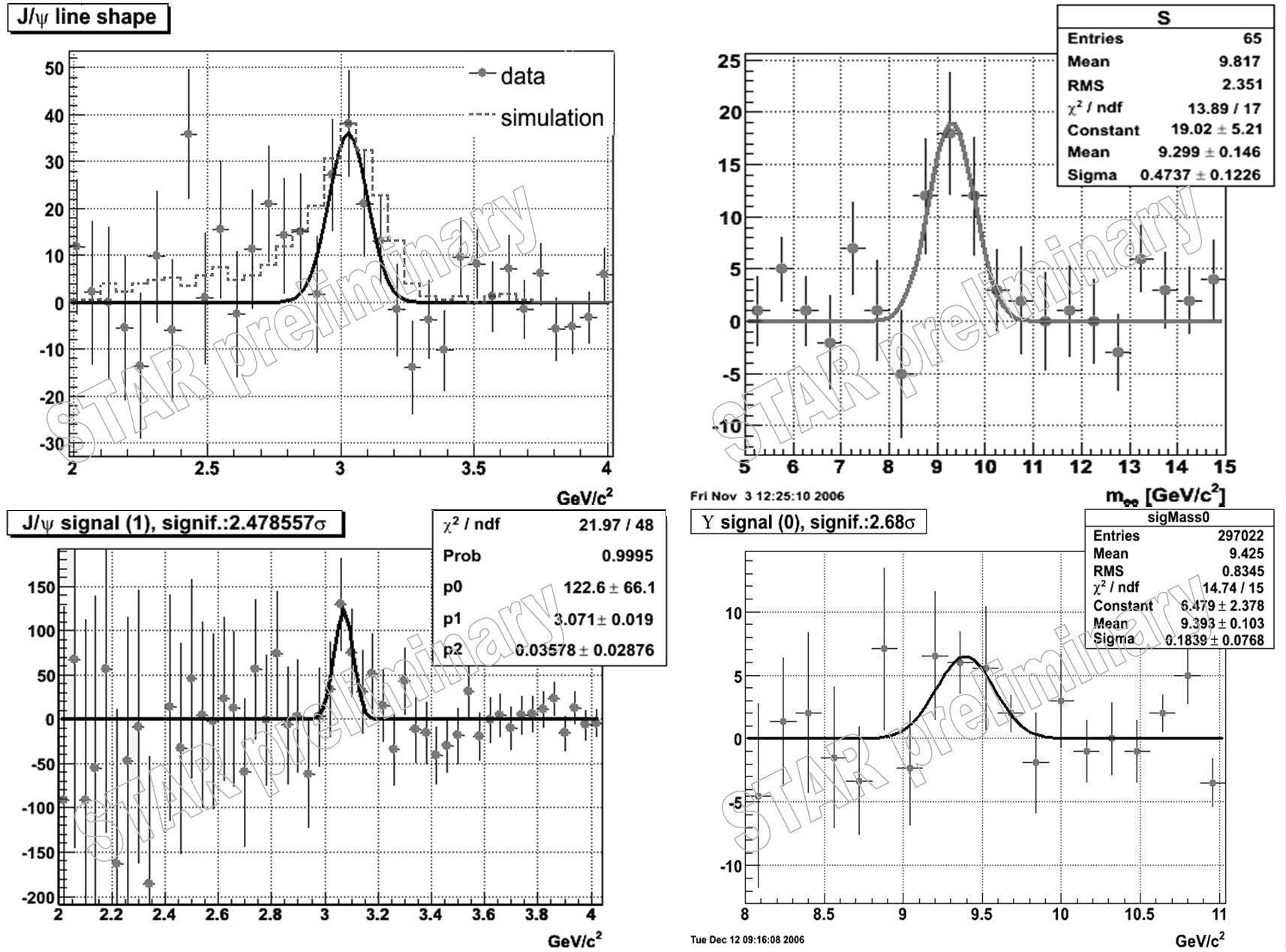}
\end{center}
\vspace{-0.8cm}
\caption{Invariant mass signals for the different quarkonia states: First row shows data from p+p, $J/\psi$ (left) and $\Upsilon$ and in bottom line same sequence for Cu+Cu data. Two important remarks on this plot: (a)$J/\psi$ data in p+p follows more closely the simulated line shape, indicating that a Gaussian fit is a poor approximation and (b) the $\Upsilon$ signal in Cu+Cu is on the limit of statistical significance to extract a cross section.}
\end{figure}

\begin{figure}[th]
\begin{center}
\includegraphics[width=1.0\textwidth]{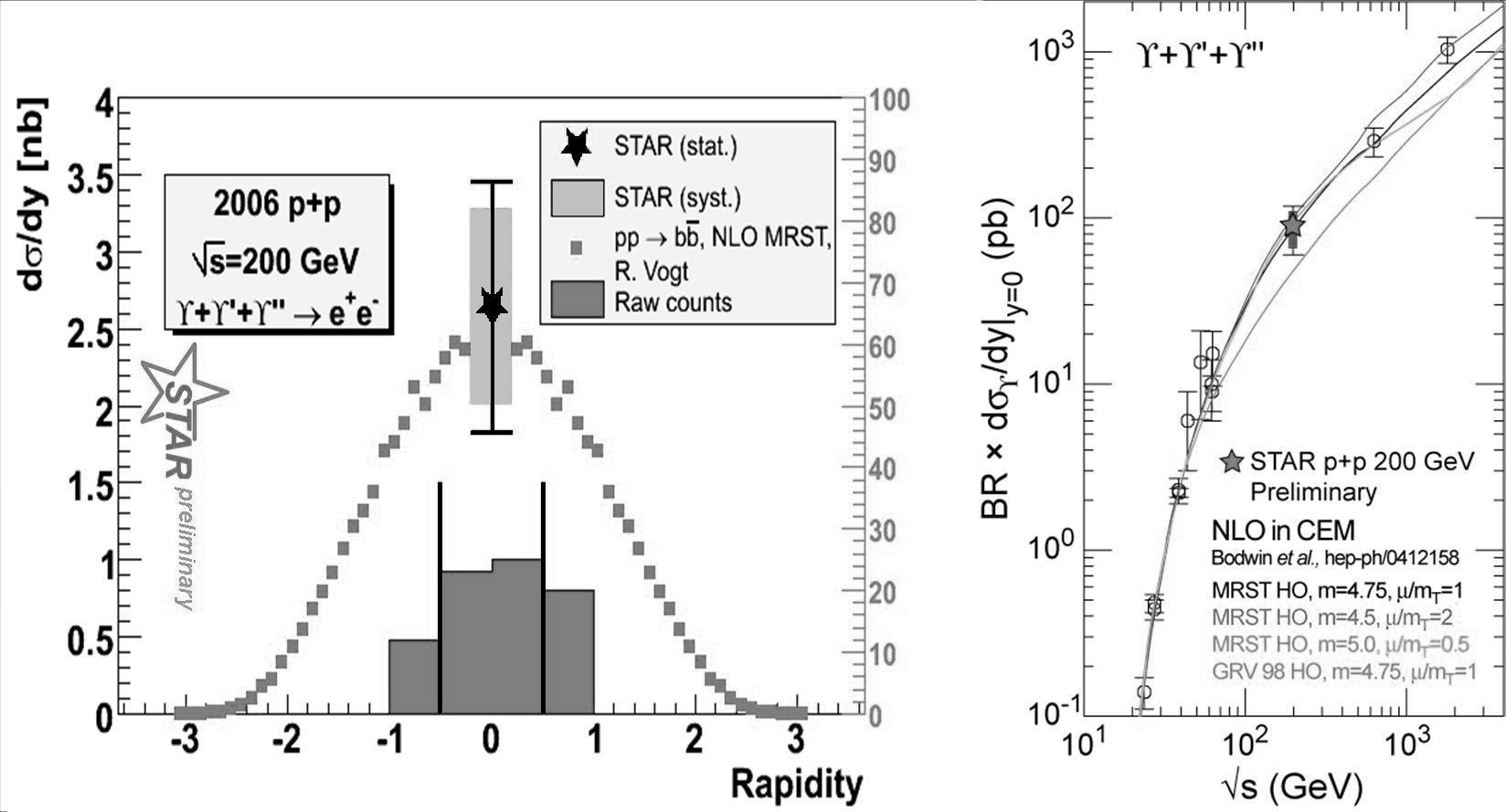}
\end{center}
\vspace{-0.8cm}
\caption{$\Upsilon$ cross section in STAR compared with theoretical values (pQCD-CEM calculation\cite{7,8}) and world data.}
\end{figure}

\section{Perspectives}
The heavy-quarkonia program in STAR will strongly benefit from the future upgrades of RHIC and STAR. The major impact 
upgrades are the luminosity enhancement ($\sim$40$\times$) on the RHIC side and, from the STAR side a barrel
Time-of-Flight detector (improves low $p_T$ PID), a barrel muon detector providing $\mu^+\mu^-$ measurements and the
DAQ1000 allowing zero dead time in special triggers for rare probes.

\section*{Acknowledgements}

This work was supported in part by the Brookhaven National Laboratory (BNL), by the 
Conselho Nacional de Desenvolvimento Cient\'{\i}fico e Tecnol\'{o}gico (CNPq) and 
by the Coordena\c{c}\~{a}o de Aperfei\c{c}oamento de Pessoal de N\'{\i}vel Superior
(CAPES).

\end{document}